\documentclass[aps,preprint,groupedaddress,showpacs,nofootinbib]{revtex4}
\usepackage{graphicx}
\newcommand{\eg}{e.g., } 
\newcommand{\ie}{i.e., } 
\newcommand{\OM}{\Omega_m} 
\newcommand{\OO}{\Omega_0}
\newcommand{\OL}{\Omega_{\Lambda}} 
\newcommand{\be}{\begin{equation}}
\newcommand{\ee}{\end{equation}}

\begin{document}
\bibliographystyle{apsrev}
\title{The Age-Redshift Relation for Standard Cosmology}
\author{R. C. Thomas \& R. Kantowski}
\email{thomas@mail.nhn.ou.edu, kantowski@mail.nhn.ou.edu}
\affiliation{
Department of Physics and Astronomy, University of Oklahoma\\
Norman, OK 73019, USA}
\date{\today}
\begin{abstract}
We present compact, analytic expressions for the 
age-redshift relation $\tau(z)$ for standard Friedmann-Lema\^\i tre-Robertson-Walker 
(FLRW) cosmology.  The new expressions are given in terms of incomplete
Legendre elliptic integrals and evaluate much faster than by direct
numerical integration.
\end{abstract}
\maketitle
\newpage
\section{Introduction}

Since Type Ia supernova observations \cite{SB,PS} have favored a Universe with $\Lambda \neq 0$, interest in FLRW
cosmologies has grown. For this family of models 
most observational relations, \eg the 
Hubble curve, are given by integral expressions; however, 
\textcite{kantk2000} recently succeeded in giving useful analytic 
distance-redshift relations 
for them. In this paper we extend these analytic results to 
include the age-redshift relation $\tau(z)$.   
These new expressions are useful 
for any computation 
that requires a transformation 
$\tau(z)$ from the observed redshift variable $z$ to the age $\tau$ of the Universe 
at that $z$. 
\textcite{FB} provides related light travel times in terms of Legendre elliptic
integrals; however, his expressions are not easy to make use of. 
A presentation closer to what we give appears in \cite{ED}. 
There, light travel time as a function of redshift was given 
for the $\OO=1$ universe, see (\ref{age_B1}). For all other cases,
\textcite{ED} was only able to give  $\tau$ and $z$ parametrically 
as Jacobi elliptic functions of conformal time $\omega\equiv \int dt/R(t)$.\footnote{
The error detected when numerical checks were performed by \textcite{CL} 
was evidently caused by an error in equation 361.54 of \textcite{BF}, 
see footnote 3 of \cite{kantk2000}}

We have concentrated on giving useful and valid expressions  for 
$\tau(z)$ appropriate for \underline{all} big bang models in the first quadrant of the
 $\OM$--\,$\OL$ plane. Because  the incomplete Legendre elliptic integrals  have 
branch points, more than one expression is necessary to completely cover this domain, 
\eg see (\ref{age_A1new}) and (\ref{age_A1}). 
 These new expressions should be quite useful for 
everything from gravitational lensing to high $z$ evolution 
studies. As an example \cite{MP,SR,RP} and \cite{DT} 
all estimate event rates of supernovae at very high ($z>1$) redshifts,
given an observed star formation rate.  Computing such event rates for any
choice of ($\Omega_m$,$\Omega_\Lambda$) requires the transformation 
$\tau(z)$.  We have made similar estimates of event rates 
and find that our computations are reduced from hours down to minutes when our
analytic $\tau(z)$ is used.\footnote{
 FORTRAN 90 and {\it 
 Mathematica} implementations of the results presented here 
 are available at http://www.nhn.ou.edu/\~{ }thomas/z2t.html.}
In \S 2 we present our results and in \S 3 some conclusions.

\section{Age of the Universe in Terms of Legendre Elliptic Integrals}

The expression for the age of the Universe at the time a source
at redshift $z$ emits light is 

\begin{equation}
 \tau(\OM,\OL;z) = \frac{1}{H_0}
  \int^\infty_{z} 
  \frac{dz}{(1+z)\sqrt{(1+z)^2(1+\Omega_m z)-z(2+z)\Omega_\Lambda}}\,,
\label{age_integral}
\end{equation}
and can easily be derived. For a Friedmann-Robertson-Walker 
(FRW) universe, \ie $\OL=0$, (\ref{age_integral}) can be integrated in terms of 
elementary functions,
\begin{equation}
\tau(\OM,\OL=0;z) = \frac{1}{H_0}\left[{\sqrt{1+\OM z}\over (1-\OM)(1+z)}-
{\OM\over (1-\OM)^{3/2}}
\sinh^{-1}\sqrt{\OM^{-1} -1\over 1+z}\right].
  \label{age_OL=0}
\end{equation}
This expression is also valid for the Einstein-de Sitter universe, 
\ie limit $\OM\rightarrow 1$, as well as when $\OM>1$.
For massless big bang models, $\OM=0$ and $0<\OL<1$, the integral is, see \cite{LK,RH}:
\begin{equation}
\tau(\OM=0,\OL;z) = \frac{1}{H_0\sqrt{\OL}}\sinh^{-1}
\left[{1\over (1+z)\sqrt{\OL^{-1}-1}}\right].
 \label{age_OM=0}
\end{equation}
When $\OL\ne 0$ and $\OM\ne 0$, (\ref{age_integral}) becomes  an 
incomplete elliptic integral and can at best be expressed as a 
combination of the three 
independent  Legendre elliptic integrals
$F(\phi,{\rm k}), E(\phi,{\rm k}),$ and $ \Pi(\phi,\alpha^2,{\rm k})$.\footnote{
Only two of the three are needed to give $\tau(z)$ and they are defined by:
$F(\phi,{\rm k})\equiv \int_0^{\phi} 1/\sqrt{1-{\rm k}^2\sin^2\phi}\ d\phi$ and
$\Pi(\phi,\alpha^2,{\rm k})\equiv \int_0^{\phi} 1/\left[(1-\alpha^2\sin^2\phi)
\sqrt{1-{\rm k}^2\sin^2\phi}\ \right]\ d\phi$.  The particular integrals needed 
can be found in \cite{BF}.} 
The form of the 
resulting expression depends on what portion of the $\OM$--\,$\OL$ plane
is being investigated. Because the cubic under the radical in (\ref{age_integral}) is 
the same as that contained in
integrals for the luminosity distance as given by \cite{kantk2000} and \cite{KR} 
a similar analysis is required. Below we outline results, 
hoping to make our expressions easy to use.

{\bf A.} $\Omega_0 = \Omega_m + \Omega_\Lambda \neq 1$

As seen below, $\tau(z)$ depends on $(\OM,\,\OL)$ primarily 
through a single  parameter $b$ 
\begin{equation}
 b \equiv -(27/2)\frac{\Omega_m^2 \Omega_\Lambda}{(1-\Omega_0)^3}.
\label{b}
\end{equation}
This parameter divides the $\OM$--\,$\OL$ plane (see Fig. 1) into
four domains where the results of integrating (\ref{age_integral})
differ. We will ignore one of the four domains and its $b=2$ boundary 
where big bangs don't occur.
In the following we use the familiar parameter  $\kappa\equiv$ ($\OO-1$)/$|\OO-1|$, 
which is determined by the sign of the 
3-curvature, to distinguish between open and closed models.
When $b<0$,  $\kappa = -1$ and when $b>0$,  $\kappa = +1$. 
Results for special boundaries other than $\OL=0$, \ie (\ref{age_OL=0}) and 
$\OM=0$, \ie (\ref{age_OM=0}),
are given in subsection {\bf B} below. The three special boundaries needed are: 
$b=\pm\infty$ \ie $\OO\equiv\OM+\OL=1$; $b=2$; and $b=27(2+\sqrt{2})/8$.

{\bf 1.} Results for the two large domains, $b < 0$ and $2 < b$, 
can be combined by defining intermediate
constants $v_\kappa$, $y_1$ and $A$:
\begin{equation}
 v_\kappa \equiv \biggl(\kappa(b-1)+\sqrt{b(b-2)}\biggr)^{1/3},
\end{equation}
\begin{equation}
 y_1 \equiv \frac{-1 + \kappa (v_\kappa+v_\kappa^{-1})}{3},
\end{equation}
\begin{equation}
 A \equiv \sqrt{y_1(3 y_1+2)}.
\label{A}
\end{equation}
These constants depend on $b$ alone and are only used for 
convenience of presentation. If the reader desires, the source of these parameters can be found in 
\cite{KR}. 
For this case we give two expressions for the integral (\ref{age_integral}).
Both are valid except for special combinations of $\OM, \OL,$ and $z$. If 
one fails the other can be used. These expressions fail when $ \Pi(\phi,\alpha^2,{\rm k})$ and 
the logarithm have canceling infinities. Both can simultaneously fail only when 
$\OM, \OL$ values are
on the $b=27(2+\sqrt{2})/8$ curve and then only for a specific value of $z$ 
(see Figs. 2 and 3). 
This special $b$ case is given in 
(\ref{age_B3}) and is good for any $z$. 
The first expression is:
\begin{eqnarray}
 \tau(\OM,\OL;z) & = & \frac{\Omega_m}{H_0|1-\Omega_0|^{3/2}} 
		\biggl[
                 \frac{1}{2\kappa y_1\sqrt{A}}\ F(\phi_z,{\rm k})
+ \frac{A-\kappa }{2 y_1 (1+y_1)\sqrt{A}}\
\Pi\biggl(\phi_z,\frac{y_1 (1+y_1)}{(A - \kappa y_1)^2},{\rm k}\biggr) \nonumber \\
           &   & 
                 + \frac{1}{2 \kappa y_1 \sqrt{\kappa (y_1+1)}}\
                 \ln\bigl(h_z^-/h_z^+\bigr)
             \biggr],
\label{age_A1new}
\end{eqnarray}
where 
\begin{eqnarray}
 h_z^{\mp}\equiv &\mp&  2\kappa y_1
\sqrt{(1+y_1)\{y_1^2(1+y_1)-[(1+z)\Omega_m/(1-\Omega_0)]^2[1+(1+z)\Omega_m/(1-\Omega_0)]\}}
\nonumber\\
&&+[(1+z)\Omega_m/(1-\Omega_0)]^2(A-\kappa y_1)-
2\kappa y_1^2(1+y_1).
\end{eqnarray}
The second expression is obtained from the first by using a ``special 
addition formula" analytically extended from a corrected version of equation 17.03 of \cite{BF}.
This transformation changes the $\alpha^2$
value of $ \Pi(\phi,\alpha^2,{\rm k})$ and hence moves the associated branch point.
The resulting second expression is: 
\begin{eqnarray}
 \tau(\OM,\OL;z) & = & \frac{\Omega_m}{H_0|1-\Omega_0|^{3/2}} 
                 \frac{1}{\sqrt{A}}
                 \biggl[
                 - \frac{F(\phi_z,{\rm k})}{A+\kappa y_1} \nonumber \\
           &   & - \frac{A-\kappa y_1}{2 \kappa y_1 (A + \kappa y_1)}
                 \Pi\biggl(\phi_z,\frac{(A+\kappa y_1)^2}{4 A \kappa y_1},{\rm k}\biggr)
                 - \frac{\sqrt{A}}{2 \kappa y_1 \sqrt{\kappa (y_1+1)}}
                 \ln\biggl(\frac{1-h_z}{1+h_z}\biggr)
             \biggr],
\label{age_A1}
\end{eqnarray}
where
\begin{equation}
 h_z \equiv \sqrt{
      \frac{(1+y_1)[y_1-(1+z)\Omega_m/(1-\Omega_0)]}
           {y_1^2+[1+(1+z)\Omega_m/(1-\Omega_0)]
            [y_1+(1+z)\Omega_m/(1-\Omega_0)]}
     }.
\end{equation}
In both cases  ${\rm k}$ and  $\phi_z$, respectively the modulus and argument of the 
elliptic integrals,
are defined by:
\begin{equation}
 {\rm k} \equiv \sqrt{\frac{2A+\kappa(1+3y_1)}{4A}},
\end{equation}
\begin{equation}
 \phi_z \equiv \cos^{-1} 
        \biggl(
         \frac{\kappa y_1+(1+z)\Omega_m/|1-\Omega_0|-A}
              {\kappa y_1+(1+z)\Omega_m/|1-\Omega_0|+A}
        \biggr).
\end{equation}

In (\ref{age_A1new}) and (\ref{age_A1}) the $z$ dependence of $\tau$ is contained 
in $\phi_z$, $h_z^{\mp}$, and $h_z$. All other terms depend on $\OM$ and $\OL$, and are easily evaluated using 
(\ref{b})-(\ref{A}). In Figures 2 and 3 the dotted lines show points where 
the first expression (\ref{age_A1new}) fails for $z = 1$ and $z = 2$ respectively.
Failure of the second expression (\ref{age_A1}) is shown by the dashed lines. Notice
that these curves always intersect somewhere on the $b=27(2+\sqrt{2})/8$ curve for a common redshift.

{\rm 2.}  If $0 < b \leq 2$, we define the three different intermediate
parameters $y_1$, $y_2$ and $y_3$
\begin{eqnarray}
 y_1 & \equiv & -\frac{1}{3}+\frac{1}{3}\cos\biggl(\frac{\cos^{-1}(1-b)}{3}\biggr)
       +\frac{1}{\sqrt{3}}\sin\biggl(\frac{\cos^{-1}(1-b)}{3}\biggr),
       \nonumber \\
 y_2 & \equiv & -\frac{1}{3}-\frac{2}{3}\cos\biggl(\frac{\cos^{-1}(1-b)}{3}\biggr),
       \nonumber \\
 y_3 & \equiv & -\frac{1}{3}+\frac{1}{3}\cos\biggl(\frac{\cos^{-1}(1-b)}{3}\biggr)
       -\frac{1}{\sqrt{3}}\sin\biggl(\frac{\cos^{-1}(1-b)}{3}\biggr).
\label{y123}
\end{eqnarray}
For this case (\ref{age_integral}) integrates to give 
\begin{equation}
 \tau(\OM,\OL;z) = \frac{\Omega_m}{H_0(\Omega_0-1)^{3/2}} 
             \frac{2}{y_1\sqrt{y_1-y_2}}
             \biggl[\Pi\biggl(\phi_z,\frac{y_1}{y_1-y_2},{\rm k}\biggr) 
             - F(\phi_z,{\rm k})\biggr],
\label{age_A2}
\end{equation}
where ${\rm k}$ and $\phi_z$ are defined by

\begin{equation}
 {\rm k} \equiv \sqrt{\frac{y_1-y_3}{y_1-y_2}},
\end{equation}
\begin{equation}
 \phi_z \equiv \sin^{-1} \sqrt{\frac{y_1-y_2}{y_1-(1+z)\Omega_m/(1-\Omega_0)}}\,.
\end{equation}
In (\ref{age_A2}) the $z$ dependence of $\tau$ is contained in $\phi_z$. 
All other terms depend on $\OM$ and $\OL$, and are easily evaluated using 
(\ref{b}) and (\ref{y123}).
There are two domains in the $\OM$--\,$\OL$ plane 
where $0 < b \leq 2$; however,
the result for this case (\ref{age_A2}) applies only to
those models which have big bangs.

{\bf B.} Special Cases

{\rm 1.} $\Omega_0 = \Omega_m + \Omega_\Lambda = 1$

This is the spatially flat model ($b\rightarrow \pm\infty$) and for it
the age-redshift integral takes on a simpler form. This
result is easily obtained using elementary integration methods. This 
result is well known:
\begin{eqnarray}
 \tau(\OM,\OL=1-\OM;z) & = & \frac{1}{H_0}
  \int^\infty_{z} 
  \frac{dz}{(1+z)\sqrt{1+\Omega_m z(3+3z+z^2)}} \nonumber \\
  & = & \frac{2}{3H_0\sqrt{1-\Omega_m}}
\ \sinh^{-1}\left(\sqrt{{\OM^{-1}-1\over (1+z)^3}}\right).
\label{age_B1}
\end{eqnarray}

{\rm 2.} $b=2$

This value of $b$ can be identified with ``critical" values of the cosmic parameters \cite{FJ}. 
The following result is equivalent to the $b=2$ value given in (\ref{age_A2}); 
however, it
is a much simpler expression:
\begin{eqnarray}
 \tau(\OM,\OL(\OM);z)  =  \frac{1}{H_0\sqrt{\OL}}
	&\ln&\ \Biggl[
	\left(
	{\sqrt{1/3-(1+z)\Omega_m/(1-\Omega_0)} +1\over
	\sqrt{1/3-(1+z)\Omega_m/(1-\Omega_0)} -1}\right)^{1/\sqrt{3}}\nonumber \\
&    & \times	\left(
	{\sqrt{1-3(1+z)\Omega_m/(1-\Omega_0)} -1\over
	\sqrt{1-3(1+z)\Omega_m/(1-\Omega_0)} +1}\right)
	\Biggr].
\label{age_B2}
\end{eqnarray}
This $\tau(z)$ doesn't apply to the Einstein-Lema\^\i tre universe 
($b=2$ where $\OL>\OM/2$) \cite{LA}, which starts expanding from the finite 
static Einstein radius 
at $t=-\infty$. However, it does apply to the $\OL<\OM/2$ models which start with a big bang 
and expand to the Einstein radius at $t=+\infty$. In the $\OM$--\,$\OL$ plane 
the static Einstein universe itself is 
a point at $\infty$ on the two $b=2$ curves where $\OM/\OL\rightarrow 2$.
If wanted, the  $b$ = constant $\ge 2$ curves can be drawn using the following
expressions. Because $\Omega_{\Lambda}(\Omega_m)$ is double valued, two expressions must be given.
For the upper part of the curve:
\be
\OO-1=3\sqrt{2/b}\ \OM\ \cosh\left[{\cosh^{-1}
\left[\sqrt{b/2}\ (\OM^{-1}-1)\right]\over 3}\right],
\ee
where 
$0\le \OM\le 1/(1-\sqrt{2/b})$.
In this expression hyperbolic cosine analytically 
becomes cosine for $\OM\ge 1/(1+\sqrt{2/b})$. 

For the lower part of the curve:
\be
\OO-1=3\sqrt{2/b}\ \OM\ \cos\left[{\cos^{-1}
\left[\sqrt{b/2}\ (1-\OM^{-1})\right]+\pi\over 3}\right],
\ee
where 
$1\le \OM\le 1/(1-\sqrt{2/b})$.
For $b=2$ (see Fig. 1) the max value of $\OM$ is `$\infty$' (the static Einstein universe);
however, for the next case (see Fig. 3) the upper and lower parts of the curve 
meet at finite $\OM\approx 1.7$.

{\rm 3.} $b=27(2+\sqrt{2})/8$

This result is equivalent to the values given by (\ref{age_A1new}) and (\ref{age_A1})
except for certain redshifts where canceling infinities appear in  
$ \Pi(\phi,\alpha^2,{\rm k})$ and the respective logarithms.
It is a simpler expression and is valid for all $z$ values, 
for this particular $b$,
\begin{eqnarray}
 \tau(\OM,\OL(\OM);z) & = & \frac{1}{H_0\sqrt{\OL}}\Bigl[
	{\sqrt{2}-1\over 4}\ F\left(\phi_z,{\sqrt{1+2\sqrt{2}}\over 2}\ \right)
	+\frac{1}{4}\ln\bigl(h_z^+/h_z^-\bigr)\Bigr],
\label{age_B3}
\end{eqnarray}
where
\begin{eqnarray}
 h_z^{\pm}&\equiv& 
	[1-(1+z)\Omega_m/(1-\Omega_0)]
	\sqrt{2[(1+z)\Omega_m/(1-\Omega_0)]^2+
	[\sqrt{2}(1+z)\Omega_m/(1-\Omega_0)+1](\sqrt{2}+1)}\nonumber\\
&&\pm (\sqrt{2}-1)[ (1+z)\Omega_m/(1-\Omega_0)+\sqrt{2}+1]
	\sqrt{(\sqrt{2}+1)[1-\sqrt{2}(1+z)\Omega_m/(1-\Omega_0)]},
\end{eqnarray}
and
\begin{equation}
 \phi_z \equiv \cos^{-1} 
        \biggl(
         \frac{-1-(1+z)\Omega_m/(1-\Omega_0)}
              {\sqrt{2}+1-(1+z)\Omega_m/(1-\Omega_0)}
        \biggr).
\end{equation}
We found it necessary to compute $\tau(z)$ for this particular $b$ value to overcome 
the occasional simultaneous failures of (\ref{age_A1new}) and (\ref{age_A1}),
see Figures 2 and 3.
\section{Conclusions}
We have given valid analytic expressions for  $\tau(z)$ in FLRW, 
the age of the Universe as a function of redshift,
which are relatively simple and are quite useful when a fast computer 
implementation is needed.\footnote{FORTRAN 90 implementation of the results 
presented here and available
at http://www.nhn.ou.edu/$\sim$thomas/z2t.html are 20-40 times faster
than a traditional Bulirsch-Stoer integrator \cite{PTVF}.}
These expressions completely cover the big bang models of the first quadrant of the 
($\OM$,\,$\OL$) plane.
If lookback times are wanted they can additionally be obtained from results given here 
by simply evaluating $\tau(0)-\tau(z)$.
Readers that are interested in adding radiation pressure as a source of
gravity should see \cite{AA} and \cite{DM} and cited references.

Even though we give several expressions for $\tau(z)$, 
most of the ($\OM$,\,$\OL$)
plane, which includes currently favored values,  is covered by case A1, 
\ie result (\ref{age_A1new}) or (\ref{age_A1}).
If $\tau(z)$ for the flat model, $\OO=1$, is wanted, the simpler result 
(\ref{age_B1}) should be used. Results for $\OO\ne 1$, (\ref{age_A1new}), 
(\ref{age_A1}), and (\ref{age_A2}), 
appear complicated 
because of the presence of extra constants, \eg $A$ and $y_1$ that 
have been retained
to compactify formulas. The reader should keep in mind that 
these are simply constants that depend on ($\OM$,\,$\OL$) through the single combination $b$
of (\ref{b}). We could have eliminated these auxiliary constants and given 
$\tau(z)$ directly in terms of the two parameters $\OM$ and $\OL$; however, 
such expressions would take up more than a page.

Expressions (\ref{age_A1new}) and  (\ref{age_A1}) for 
$\tau(z)$ remain real but as presented can contain imaginary terms because of branch points. 
The threshold is defined by $1-\alpha^2\sin^2\phi_z=0$ in $\Pi(\phi_z,\alpha^2,{\rm k})$. 
If $1-\alpha^2\sin^2\phi_z<0$ canceling imaginary terms appear in $\Pi(\phi_z,\alpha^2,{\rm k})$ 
and the logarithm. For expressions that avoid this imaginary complication the 
reader simply replaces the argument of the logarithm with its magnitude and 
$\Pi(\phi_z,\alpha^2,{\rm k})$ with its principal part. At threshold points where 
$1-\alpha^2\sin^2\phi_z=0$, canceling infinities appear in $\Pi(\phi_z,\alpha^2,{\rm k})$
and the logarithm. The infinity problem is avoided by switching between (\ref{age_A1new}) and  
(\ref{age_A1}). If both have infinities then (\ref{age_B3}) gives the correct result. 
\acknowledgements
The authors thank P. Helbig for input and J. E. Felten 
for suggesting additional references.

\clearpage
\begin{figure}
\includegraphics{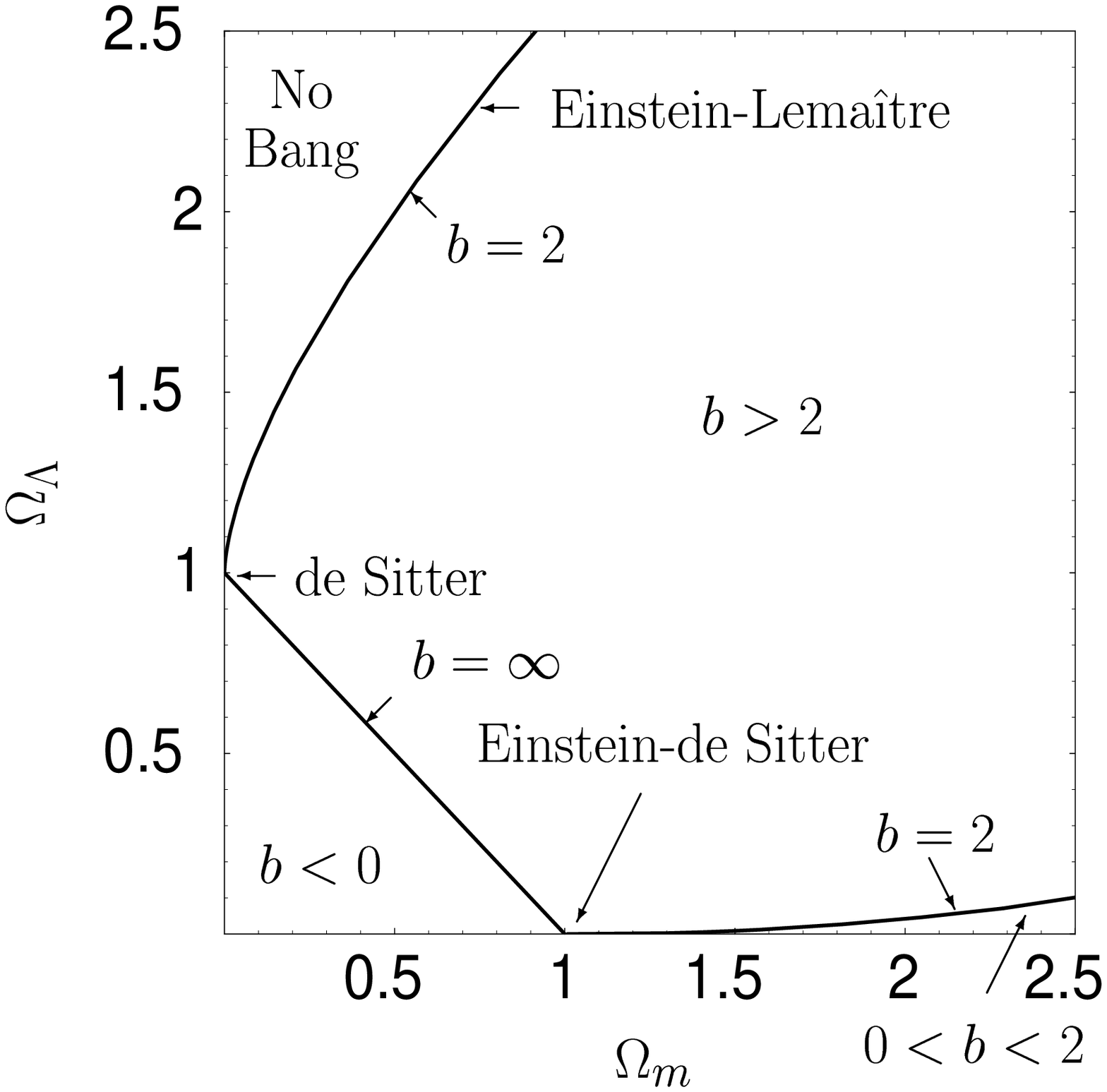}
\caption[fig1_3.eps]{The $\OM$--\,$\OL$ plane showing various $b$ domains 
that require different expressions for age-redshift $\tau(z)$ for
standard FLRW. Expressions (\ref{age_A1new}) and (\ref{age_A1})
are both appropriate for $b<0$ and $b>2$, while (\ref{age_B1}) is appropriate for 
$0<b\le 2$. Simpler expressions exist for various boundaries: $\OL=0$ (\ref{age_OL=0}), $\OM=0$ (\ref{age_OM=0}),
$b\rightarrow\infty \Leftrightarrow \OO=1$ (\ref{age_B1}), and  $b=2$ (\ref{age_B2}).
\label{fig1_3}}
\end{figure}
\begin{figure}
\includegraphics{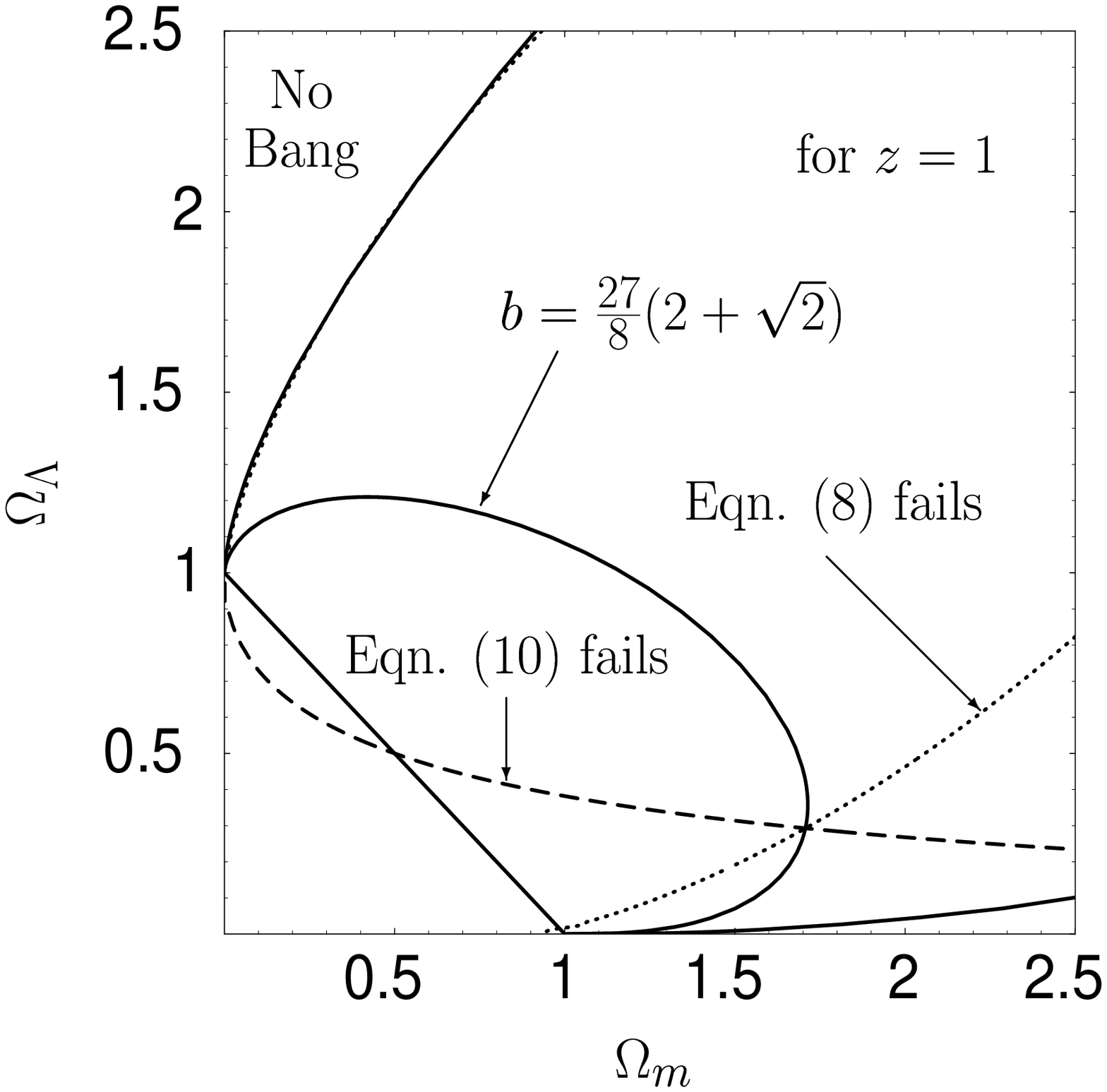}
\caption[fig2_3.eps]{The $\OM$--\,$\OL$ plane showing 
curves where  age-redshift $\tau(z)$ expressions (\ref{age_A1new}) and (\ref{age_A1})
fail for redshift $z=1$. Equation (\ref{age_A1new}) fails along the dotted curve
and (\ref{age_A1}) fails along the dashed curve. Both fail where they intersect
on the $b=27(2+\sqrt{2})/8$ curve; however, (\ref{age_B3}) gives the $\tau(z)$ 
value at any point on this curve for all redshifts.
\label{fig2_3}}
\end{figure}

\begin{figure}
\includegraphics{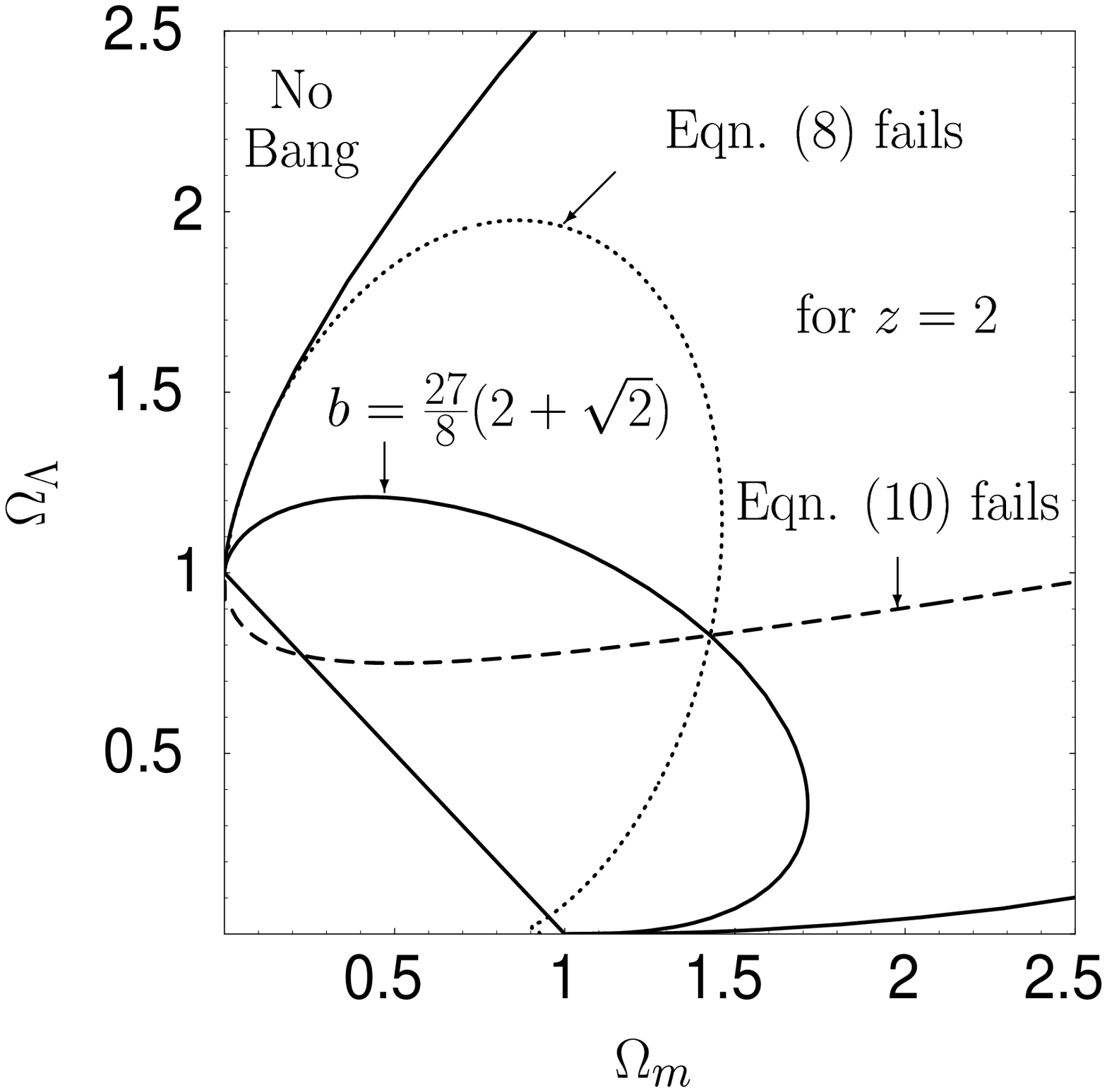}
\caption[fig3_3.eps]{Same as Fig. 2 except for redshift $z=2$.
\label{fig3_3}}
\end{figure}


\begin{thebibliography}{}

\bibitem[Agnese et al.(1970)]{AA}A. Agnese, M. La  Camera  and A. Wataghin,
Il Nuovo Cimento {\bf 66}, 202 (1970).

\bibitem[Byrd \& Friedman(1954-71)]{BF}P. F. Byrd and M. D. Friedman,
{\it Handbook of Elliptic Integrals for Engineers \& Scientists} 
(Springer-Verlag, New York, 1954-71).

\bibitem[Campusano et al.(1975)]{CL}L. Campusano, J. Heidmann, and J. L. Nieto, A. \& A. {\bf 41}, 229 (1975).

\bibitem[Dabrowski \& Stelmach(1986)]{DM}M. Dabrowski and J. Stelmach, A. J. {\bf 92}, 1272 (1986).

\bibitem[Dahlen \& Fransson(1999)]{DT}T. Dahlen and C. Fransson, A. \& A. {\bf 350}, 349 (1999).

\bibitem[Edwards(1972)]{ED}D. Edwards, M.N.R.A.S. {\bf 159}, 51 (1972).

\bibitem[Feige(1992)]{FB}B. Feige, Astron. Nachr. {\bf 313}, 139 (1992).

\bibitem[Felten(1986)]{FJ}J. E. Felten and R. Isaacman, Rev. Mod. Phys. {\bf 58}, 689 (1986).

\bibitem[Kantowski et al.(2000)]{kantk2000}R. Kantowski, J. K. Kao, and R. C. Thomas, astro-ph/0002334, in press Ap. J. {\bf 545}, Dec. 10 (2000). 

\bibitem[Kantowski(1998)]{KR}R. Kantowski, Ap. J. {\bf 507}, 483 (1998).

\bibitem[Lanczos(1922)]{LK}K. Lanczos, Phys. Zeit. {\bf 23}, 539 (1922).

\bibitem[Lema\^\i tre(1931)]{LA}A. G. Lema\^\i tre, M.N.R.A.S. {\bf 91}, 483 (1931).

\bibitem[Madau et al.(1998)]{MP}P. Madau, M. Della Valle, and N. Panagia,
M.N.R.A.S. {\bf 297}, L17 (1998).

\bibitem[Perlmutter et al.(1998)]{PS}S. Perlmutter et al., Ap. J. {\bf 517}, 565 (1999).

\bibitem[Press et al.(1994)]{PTVF}W. Press, S. Teukolsky, W. Vetterling, and B. Flannery, 
{\it Numerical Recipes} (Cambridge University Press, Cambridge, 1994). 

\bibitem[Robertson(1933)]{RH}H. P. Robertson, Rev. Mod. Phys. {\bf 5}, 62 (1933).

\bibitem[Ruiz-Lapuente \& Canal(1998)]{RP}P. Ruiz-Lapuente and R. Canal, Ap. J. {\bf 497}, L57 (1998).

\bibitem[Sadat et al.(1998)]{SR}R. Sadat, A. Blanchard, B. Guideroni, and J. Silk,
 A. \& A. {\bf 331}, L69 (1998).

\bibitem[Schmidt et al.(1998)]{SB}B. P. Schmidt et al., Ap. J. {\bf 507}, 46 (1998). 

\end{thebibliography}
\end{document}